\documentclass[aps,twocolumn,superscriptaddress,showpacs]{revtex4}

\usepackage[dvips]{graphicx}
\usepackage{color}  % for color  usage: \textcolor{red}{text}
\usepackage{amssymb}
\newcommand{\nc}{\newcommand}           % new command
\nc{\vc}[1]     {\mbox{\boldmath $#1$}} % boldmath(vector)
\nc{\mapleft}[1]{                       % something under arrow
 \smash{\mathop{                      %
  \hbox to 0.90cm{\rightarrowfill} }\limits_{#1}}}

\nc{\beq}     {\begin{eqnarray}}
\nc{\eeq}    {\end{eqnarray}}

\nc{\mydraft}	{\setlength{\topmargin}{-1.5cm}}
\mydraft
\begin{document}

\title{
Power series expansion method in tensor--optimized antisymmetrized\\ molecular dynamics beyond the Jastrow correlation method
}

\author{Takayuki Myo\footnote{takayuki.myo@oit.ac.jp}}
\affiliation{General Education, Faculty of Engineering, Osaka Institute of Technology, Osaka, Osaka 535-8585, Japan}
\affiliation{Research Center for Nuclear Physics (RCNP), Osaka University, Ibaraki, Osaka 567-0047, Japan}

\author{Hiroshi Toki\footnote{toki@rcnp.osaka-u.ac.jp}}
\affiliation{Research Center for Nuclear Physics (RCNP), Osaka University, Ibaraki, Osaka 567-0047, Japan}

\author{Kiyomi Ikeda\footnote{k-ikeda@postman.riken.go.jp}}
\affiliation{RIKEN Nishina Center, Wako, Saitama 351-0198, Japan}

\author{Hisashi Horiuchi\footnote{horiuchi@rcnp.osaka-u.ac.jp}}
\affiliation{Research Center for Nuclear Physics (RCNP), Osaka University, Ibaraki, Osaka 567-0047, Japan}

\author{Tadahiro Suhara\footnote{suhara@matsue-ct.ac.jp}}
\affiliation{Matsue College of Technology, Matsue 690-8518, Japan}

\date{\today}

\begin{abstract}%
We developed a new variational method for tensor-optimized antisymmetrized molecular dynamics (TOAMD) for nuclei. 
In TOAMD, the correlation functions for the tensor force and the short-range repulsion are introduced and used in the power series form of the wave function, which is different from the Jastrow method. 
Here, nucleon pairs are correlated in multi-steps with different forms, while they are correlated only once including all pairs in the Jastrow correlation method. 
Each correlation function in every term is independently optimized in the variation of total energy in TOAMD. 
For $s$-shell nuclei using the nucleon-nucleon interaction, the energies in TOAMD are better than those in the variational Monte Carlo method with the Jastrow correlation function. 
This means that the power series expansion using the correlation functions in TOAMD describes the nuclei better than the Jastrow correlation method.
\end{abstract}

\maketitle

%%%%%%%%%%%%%%%%%%%%%%%%%%%%%
\section{Introduction}
The nucleon--nucleon ($NN$) interaction has a strong repulsion at a short distance and a strong tensor force at long and intermediate distances \cite{pieper01,wiringa84}.
In particular, the main component of the tensor force arises from the one--pion--exchange interaction and is the dominant force of attraction in strongly interacting systems.
These characteristics of $NN$ interaction induce short-range and tensor correlations in the nuclei.
The short-range repulsion reduces the amplitudes of nucleon pairs at the short distances in the nuclei and
the tensor force produces strong $S$--$D$ coupling, which leads to the admixture of the spatially compact $D$-wave state of nucleon pairs in the nuclei.

Jastrow introduced the method of correlation function in many-body problems \cite{jastrow55}. Since then, the Jastrow correlation method was used in essentially all the microscopic many-body calculations for the treatment of many-body problems in any discipline.
In nuclear physics, the Jastrow correlation method has become the familiar method for treating the tensor and short-range correlations in nuclei starting from the $NN$ interaction \cite{carlson15}.
In this method, the Jastrow factor, which has a product-type form of the pair correlation functions, is multiplied to the uncorrelated nuclear wave function and as a result the correlation acts only once in each particle pair.  The Jastrow correlation method is often used not only in nuclear physics \cite{carlson15}, but also the condensed matter \cite{haldane88}, and atomic and molecular physics \cite{umrigar88,casula03} to describe the central correlations.
In numerical calculations, the Monte Carlo technique, which is known as the variational Monte Carlo method (VMC), is often employed for the integration of the matrix elements.
In this method, one essentially assumes the common functional forms of correlation functions for every pair with a state dependence.
These forms are often given {\it a priori} under the appropriate asymptotic condition of pairs before the variation of the total wave function.
In the case of $NN$ interaction of nuclei, one must consider the tensor correlation as well as the short-range correlation,
which has a long-range nature and a characteristic feature of nuclei resulting in large binding energy \cite{pieper01}.

Recently, we developed a new variational method for treating the characteristics of $NN$ interaction directly \cite{myo15,myo17a,myo17b,myo17c}.
We employed antisymmetrized molecular dynamics (AMD) \cite{kanada03,kanada12} as the basis state for nuclei,
and introduced two kinds of correlation functions: the tensor-operator type and central-operator type.
The AMD wave function is suitable for describing nuclear clustering, which is one of the important aspects of nuclear structure such as the triple-$\alpha$ Hoyle state in $^{12}$C \cite{ikeda68,horiuchi12}.
The correlation functions are multiplied to the AMD wave function and these components are superposed with the original AMD wave function.
We named this method tensor--optimized antisymmetrized molecular dynamics (TOAMD) \cite{myo15}.
The concept of TOAMD originated from the tensor--optimized shell model (TOSM) \cite{myo05,myo09,myo11} in which the two-particle two-hole (2p-2h) excitations are fully optimized
in treating the tensor and short-range correlations induced by the $NN$ interaction.  
In TOAMD, the 2p-2h excitations are expressed by the tensor and short-range correlation functions and as a natural extension we introduce multiple products of the correlation functions for the improvement of the variational wave function. 

In TOAMD, the total wave function has the form of a power series expansion using the tensor and short-range correlation functions, 
where the functional forms of each correlation function are determined independently in the energy variation of the total system.
On the other hand, the Jastrow correlation function is expressed as a multiple of the pair correlation, which makes only a single correlation function in each pair in the many-body wave function.

In TOAMD, there are multiple products consisting of the Hamiltonian and correlation functions, which are expanded into a series of many-body operators using the cluster expansion \cite{myo15}.
All the resulting many-body operators are considered in the calculation to establish the TOAMD variational method.
In our previous work \cite{myo17a}, we described the $s$-shell nuclei in TOAMD with up to the double products of the correlation functions.
It was shown that TOAMD satisfactorily reproduces the results of Green's function Monte Carlo (GFMC) using the bare $NN$ interaction \cite{kamada01}.  
The TOAMD scheme can be extended by successively increasing the power series of the multiple products of the correlation functions.

The objective of this paper is to compare TOAMD and the Jastrow correlation method as the variational models using the common $NN$ interactions.
As the $NN$ interaction we used the Malfliet-Tjon interaction with only short-range repulsion and the Argonne V6 (AV6) bare $NN$ interaction with tensor force, 
for which the VMC results were reported using the Jastrow correlation method with the central-type and tensor-type two-body correlation functions.
In addition, we discuss the effect of successive and independent optimization of the correlation functions on the solutions of TOAMD.

%%%%%%%%%%%%%%%%%%%%%%%%%%%%%
\section{Tensor optimized antisymmetrized molecular dynamics (TOAMD)}\label{sec:method}
The basic framework of TOAMD is as follows, while the details of TOAMD are given in Refs. \cite{myo15,myo17c}.
The AMD wave function $\Phi_{\rm AMD}$, which is a single Slater determinant of the Gaussian wave packets of nucleons with mass number $A$, is given as:
\begin{eqnarray}
\Phi_{\rm AMD}
&=& \frac{1}{\sqrt{A!}}\, {\rm det} \left\{ \prod_{i=1}^A \phi_i \right\}\,,
\label{eq:AMD}
\\
\phi(\vec r)&=&\left(\frac{2\nu}{\pi}\right)^{3/4} e^{-\nu(\vec r-\vec D)^2}\chi_{\sigma}\chi_{\tau}\,.
\label{eq:Gauss}
\end{eqnarray}
The wave function of a single nucleon $\phi(\vec r)$ consists of a Gaussian wave packet with a range parameter $\nu$, centroid position $\vec D$, 
spin part $\chi_{\sigma}$, and isospin part $\chi_{\tau}$.
In this work, $\chi_{\sigma}$ is the up or down component and $\chi_{\tau}$ is a proton or neutron.
The range parameter $\nu$ is common for all the nucleons. This condition eliminates the center-of-mass excitation in both AMD and TOAMD.
The AMD wave function can be extended to include multi-configurations applying the generator coordinate method using various centroid positions $\vec D$, 
although we employ the single configuration of AMD in the present study.

We introduce two kinds of pair correlation functions: $F_D$ for tensor force and $F_S$ for short-range central repulsion.
The explicit forms of the correlation functions including state dependence are given as
\begin{eqnarray}
F_D
&=& \sum_{t=0}^1\sum_{i<j}^A f^{t}_{D}(r_{ij})\,r_{ij}^2\, S_{12}(\hat{r}_{ij})\,(\vec \tau_i\cdot \vec \tau_j)^t \,,
\label{eq:Fd}
\\
F_S
&=& \sum_{t=0}^1\sum_{s=0}^1\sum_{i<j}^A f^{t,s}_{S}(r_{ij})\,(\vec \tau_i\cdot \vec \tau_j)^t\,(\vec\sigma_i\cdot \vec\sigma_j)^s\,,
\label{eq:Fs}
\end{eqnarray}
with a relative distance $r_{ij}=|\vec r_i - \vec r_j|$ and tensor operator $S_{12}(\hat r)$. 
The labels $s$ and $t$ are the spin and isospin channels of the two nucleons, respectively.
The pair functions $f^{t}_{D}(r)$ and $f^{t,s}_{S}(r)$ are the variational functions.
The functions $F_D$ and $F_S$ can excite two nucleons to the high-momentum state in nuclei, which corresponds to the 2p-2h excitations.
It should be noted that $F_D$ and $F_S$ can also express the long-range correlation.

These correlation functions are multiplied to the AMD wave function $\Phi_{\rm AMD}$ and the resulting components are superposed with $\Phi_{\rm AMD}$.
We further take the power series expansion in terms of the multiple correlations consisting of $F_D$ and $F_S$.
In the present work, we consider up to the double products of the correlation functions, which can represent up to the 4p-4h excitations.
The TOAMD wave function is defined as
\begin{eqnarray}
\Phi_{\rm TOAMD}
&=& (1+F_S+F_D+F_S F_S+F_S F_D
\nonumber\\
&& +\,F_D F_S+F_D F_D) \times\Phi_{\rm AMD}~.
\label{eq:TOAMD2}
\end{eqnarray}
It should be noted that $F_D$ and $F_S$ in each term of Eq. (\ref{eq:TOAMD2}) are independent.
Hence, there are five kinds of each of $F_D$ and $F_S$, which are fully optimized for the energy variation.
For simplicity, we use the common symbols $F_D$ and $F_S$.
This expression of the TOAMD wave function is commonly used for all nuclei.

The relation between TOAMD and the Jastrow correlation method \cite{myo17a} is briefly explained.
In the Jastrow correlation method, the product-type correlation function is assumed to be
\begin{eqnarray}
 F_{\rm Jastrow} &=& \prod_{i<j}^{A} f_{ij}(\vec r_{ij}),
\label{eq:Jastrow}
\end{eqnarray}
where the symmetrization of the Jastrow factor $F_{\rm Jastrow}$ is omitted to simplify the discussion.
The pair function $f_{\rm}(\vec r)$ consists of the terms involving the state dependence of the spin-isospin as well as the operator dependence such as $S_{12}(\hat r)$. 
When we express the function $f(\vec r)$ as $1+\widetilde{f}(\vec r)$, the factor $F_{\rm Jastrow}$ can be expanded into the power series form using $\widetilde{f}(\vec r)$ 
in an almost similar form as that of Eq.~(\ref{eq:TOAMD2}).
The function $\widetilde{f}(\vec r)$ corresponds to the pair functions $f^{t}_{D}(r)$ and $f^{t,s}_{S}(r)$ in TOAMD.
It should be noted that $\widetilde{f}(\vec r)$ acts only once in every pair, while in TOAMD the nucleon pairs are correlated in multi-steps with different pair functions.
We should note, however, that the TOAMD wave function does not fulfill the cluster decomposition property; all the nucleon pairs are not always correlated via the correlation functions, while the Jastrow method possesses this property. In the TOAMD, we can have this property by including more correlation functions.

We use the Hamiltonian with a two-body $NN$ interaction $V$ for mass number $A$ as
\begin{eqnarray}
    H
&=& T+V
~=~ \sum_i^A t_i-T_{\rm c.m.} + \sum_{i<j}^{A} v_{ij}\,,
    \label{eq:Ham}
    \\
    v_{ij}
&=& v_{ij}^{\rm C} + v_{ij}^{\rm T}\,.
\end{eqnarray}
Here, $t_i$ and $T_{\rm c.m.}$ are the kinetic energies of each nucleon and the center-of-mass, respectively.
In the present study, we use a two-body bare $NN$ interaction AV6 consisting of central and tensor forces originating from the bare AV14 potential \cite{wiringa84}, 
which is used in the VMC, GFMC, and few-body calculations \cite{carlson87,varga97} and then suitable for comparison of the TOAMD results with other calculations.

The total energy $E$ in TOAMD is given by:
\begin{eqnarray}
    E
&=&\frac{\langle \Phi_{\rm TOAMD} |H|\Phi_{\rm TOAMD} \rangle}{\langle \Phi_{\rm TOAMD} |\Phi_{\rm TOAMD} \rangle}
    \nonumber\\
&=& \frac{\langle \Phi_{\rm AMD} |\tilde{H}|\Phi_{\rm AMD} \rangle}{\langle \Phi_{\rm AMD} | \tilde{N} |\Phi_{\rm AMD} \rangle}.
\label{eq:E_TOAMD}
\end{eqnarray}
The operators $\tilde{H}$ and $\tilde{N}$ are the correlated Hamiltonian and norm operators, respectively.
The matrix elements of the correlated operators are calculated using the AMD wave function.
The correlated operators $\tilde{H}$ and $\tilde{N}$ include various products of correlation functions such as $F^\dagger H F$ and $F^\dagger F$,
where $F$ stands for $F_D$ and $F_S$. 
These operators are expanded into the series of many-body operators in terms of the cluster expansion \cite{myo15}.
For the two-body interaction $V$, the correlated interaction $F^\dagger V F$ gives up to six-body operator.  In the same way, $F^\dagger F^\dagger V F F$ induces up to 10-body operators.

We employ all the resulting many-body operators of $\tilde{H}$ and $\tilde{N}$ without any truncation, which is necessary to retain the variational principle for TOAMD.  In general, multiple products of many correlation functions produce a large number of many-body operators in the cluster expansion.
Among these operators, the higher-body terms require larger calculation costs to obtain their matrix elements numerically, which occurs often for larger mass nuclei.

The present TOAMD wave function has two-kinds of variational functions, $\Phi_{\rm AMD}$, and $F_D$ and $F_S$.
They are determined to minimize the total energy $E$ as $\delta E=0$.
For $\Phi_{\rm AMD}$, the centroid positions $\{\vec D_i\}$ ($i=1,\cdots,A$) in Eq.~(\ref{eq:Gauss}) are determined by using the cooling method \cite{kanada03}.
We optimize the radial forms of the pair functions $f^{t}_{D}(r)$ in Eq.~(\ref{eq:Fd}) and $f^{t,s}_{S}(r)$ in Eq.~(\ref{eq:Fs}) using the Gaussian expansion method as
\begin{eqnarray}
   f^t_D(r)
&=&\sum_{n=1}^{N_G} C^t_n e^{-a^t_n\, r^2},
   \label{eq:cr_D}
   \\
   f^{t,s}_S(r)
&=&\sum_{n=1}^{N_G} C^{t,s}_n e^{-a^{t,s}_n\, r^2}.
   \label{eq:cr_S}
\end{eqnarray}
We take the Gaussian number $N_G=7$ to obtain the converging results.
We search for the values of $a^t_n$, $a^{t,s}_n$ in a wide range to cover the spatial correlation.
We determine the expansion coefficients $C^t_n$ and $C^{t,s}_n$ by the diagonalization of the Hamiltonian matrix.

In the calculation, we express the TOAMD wave function in the form of the linear combination of the basis states using the coefficients of the Gaussian functions in the correlation functions.
\begin{eqnarray}
   \Phi_{\rm TOAMD}
&=& \sum_{\alpha=0} \tilde{C}_\alpha \,\Phi_{{\rm TOAMD}, \alpha}\,,
   \label{eq:linear}
   \\
   H_{\alpha \beta}
&=& \langle \Phi_{\rm AMD}|\tilde{H}_{\alpha \beta}|\Phi_{\rm AMD} \rangle\,,
   \nonumber\\
   N_{\alpha \beta}
&=& \langle \Phi_{\rm AMD}|\tilde{N}_{\alpha \beta}|\Phi_{\rm AMD} \rangle\,,
   \nonumber
   \label{eq:HN}
\end{eqnarray}
where the labels $\alpha$ and $\beta$ represent the set of the Gaussian index $n$, quantum numbers of spin $s$ and isospin $t$ for two nucleons in the correlation functions.
The Hamiltonian and norm matrix elements are given as $H_{\alpha \beta}$ and $N_{\alpha \beta}$, respectively.
We assign the AMD wave function $\Phi_{\rm AMD}$ to the labels $\alpha=\beta=0$, and then $\tilde{H}_{0 0}=H$ and $\tilde{N}_{0 0}=1$. 
The corresponding expansion coefficient is $\tilde C_0$ in Eq.~(\ref{eq:linear}).
For the basis states with single correlation functions, $\tilde C_\alpha$ becomes $C_n^t$ and $C_n^{t,s}$ given in Eqs.~(\ref{eq:cr_D}) and (\ref{eq:cr_S}), respectively.
In the double products of the correlation functions, the products of two Gaussian functions in Eqs.~(\ref{eq:cr_D}) and (\ref{eq:cr_S}) are treated as the basis functions. 
Accordingly, the expansion coefficients $\tilde{C}_\alpha$ can be $C_n^t C_{n'}^{t'}$, $C_n^t C_{n'}^{t',s}$ and $C_n^{t,s} C_{n'}^{t',s'}$,
where the label $\alpha$ includes the information on two-kinds of the correlation functions.
The coefficients $\tilde{C}_\alpha$ cannot be decomposed into $C_n^t$ and $C_n^{t,s}$ inversely.
Finally, we solve the following eigenvalue problem and obtain the total energy $E$ and all the coefficients $\tilde{C}_\alpha$.
\begin{eqnarray}
   \sum_{\beta=0}\left(H_{\alpha\beta}-EN_{\alpha\beta} \right)\tilde{C}_\beta&=&0.
   \label{eq:eigen}
\end{eqnarray}
The explicit forms of the pair functions $f^t_D(r)$ and $f^{t,s}_S(r)$ obtained in the calculations are shown in Ref. \cite{myo17c} for $^3$H and $^4$He in the case of the AV8$^\prime$ $NN$ interaction.
Their radial behaviors are reasonable to describe the tensor and short-range correlations.

We briefly explain the procedure to calculate the matrix elements of the correlated Hamiltonian using the AMD wave function in Eq.~(\ref{eq:E_TOAMD}).
We express the $NN$ interaction $V$ using a sum of Gaussian functions in the same way as the correlation function $F$.
The correlated operators $\tilde{H}$ and $\tilde{N}$ become the products of $F$ and $V$. 
After the cluster expansion of $\tilde{H}$ and $\tilde{N}$ into the independent many-body operators,
the resulting operators become the products of Gaussian functions and involve various combinations of the interparticle coordinates. 
This structure for the coordinates makes it difficult to calculate the matrix elements of many-body operators in general.
In the TOAMD, we perform the Fourier transformation of the Gaussian functions in $F$ and $V$.
In this transformation, we can decompose the square of the interparticle coordinates $\vec r_{ij}^{\,2}$ in the Gaussian functions into the plane waves having each particle coordinate $\vec r_i$ and $\vec r_j$.
Hence, in the momentum space, the matrix elements of many-body operators become the products of the single-particle matrix elements of the plane waves.
Using the single-particle matrix elements with AMD, we perform the multiple integration of the associated momenta and obtain the matrix elements of TOAMD.
We explain the above procedure in detail with typical examples in Refs.~\cite{myo15,myo17c}.

%%%%%%%%%%%%%%%%%%%%%%%%%%%%%%%%%%%%%%%%%%%%
\section{Results}\label{sec:results}
First, we discuss the comparison between TOAMD and VMC in the case of central interaction with the short-range repulsion, which is reported in Ref. \cite{myo17a} for $s$-shell nuclei in detail.
We use only the central correlation functions $F_S$ and $F_S F_S$ in Eq. (\ref{eq:TOAMD2}).
Hence, we have two kinds of correlation functions.
We chose the Malfliet-Tjon V (MT-V) $NN$ potential with strong short-range repulsion \cite{malfliet69,zabolitzky82}.
In Table \ref{tab:MTV}, we list the results of TOAMD with other wave functions.
The energies of $^3$H and $^4$He in TOAMD reproduce the results of few-body calculations satisfactorily.
Furthermore, the energies in TOAMD become lower than those of VMC using Jastrow correlation method \cite{carlson81}.
This result indicates that the variational accuracy of TOAMD is better than that of the Jastrow correlation method for the description of short-range correlation. 

%%%%%%%%%%%%%%%%%%%%%%%%%%%%%% 
% Src2.6/Data_MTV
\begin{table}[t]
\begin{center}
\caption{Energies of $^3$H($\frac12^+$) and $^4$He ($0^+$) using MT-V potential in units of MeV in comparison with other wave functions.}
\label{tab:MTV} 
\begin{tabular}{c|rrr}
\noalign{\hrule height 0.5pt}
        &~~VMC \cite{carlson81} &~~Few-body \cite{varga95} &~~~TOAMD  \\
\noalign{\hrule height 0.5pt}
$^3$H~~ &~~$-8.22(2)$~~         &~~$-8.25$~~               & ~$-8.24$~~\\
$^4$He~~&~$-31.19(5)$~~         &~$-31.36$~~               & $-31.28$~~\\
\noalign{\hrule height 0.5pt}
\end{tabular}
\end{center}
\end{table}
%%%%%%%%%%%%%%%%%%%%%%%%%%%%%%

%%%%%%%%%%%%%%%%%%%%%%%%%%%%%% 
% Src2.4/Data03_AV6/ (N_G=7)
\begin{table}[t]
\begin{center}
\caption{Energies of $^3$H ($\frac12^+$) and $^4$He ($0^+$) using AV6 potential 
by adding each correlation term in TOAMD successively in units of MeV.}
\label{tab:AV6} 
\begin{tabular}{crrrrrr}
\noalign{\hrule height 0.5pt}
        &~~AMD    &~~+S      &~~+D        &~~+SS     &~~+DS     &~~+DD \\ 
\noalign{\hrule height 0.5pt}
$^3$H   & $15.18$  & $3.48$   & $-4.57$   &  $-5.52$ & $-6.56$  & $-7.10$ \\
$^4$He  & $65.46$  & $11.17$  & $-13.87$  & $-17.15$ & $-21.20$ & $-24.31$ \\
\noalign{\hrule height 0.5pt}
\end{tabular}
\end{center}
\end{table}
%%%%%%%%%%%%%%%%%%%%%%%%%%%%%%

%%%%%%%%%%%%%%%%%%%%%%%%%%%%%% 
% Src2.9.3/Data03_AV6
\begin{table}[t]
\begin{center}
\caption{Energies of $^3$H($\frac12^+$) and $^4$He ($0^+$) using AV6 potential in units of MeV in comparison with other wave functions. The symbols $K$ and $V$ represent the kinetic and interaction energies, respectively, and C and T represent the contributions of central and tensor forces, respectively.  The units of radius is fm.}
\label{tab:AV6_2} 
\begin{tabular}{cc|rrrrrr}
\noalign{\hrule height 0.5pt}
      &      &~VMC~              &~GFMC~             &~Few-body~      &~TOAMD~   \\
      &      &\cite{carlson87}~~~&\cite{carlson87}~~~&\cite{varga97}~~~&          \\
\noalign{\hrule height 0.5pt}
      & $E$  &  $-6.33(5)$       & $-7.22(12)$       &  $-7.15$      &~$-7.10$   \\
      & $K$  &  $ 37.4$          & $44.8$            &  $44.8$       &~$44.56$   \\
$^3$H & $V$  &  $-43.7$          & $-52.0$           &  $-51.9$      &~$-51.66$  \\
      &  C   &                   &                   &               &~$-17.68$  \\ 
      &  T   &                   &                   &               &~$-33.98$  \\ 
      &Radius&  1.95(3)          & 1.75(10)          &  1.76         &~1.76      \\ \hline
      & $E$  & $-22.75(10)$      & $-24.79(20)$      & $-25.40$      & $-24.31$  \\
      & $K$  & $ 99.3$           & $97.2$            &  $100.1$      &~$ 95.17$  \\
$^4$He& $V$  & $-122 $           & $-122$            &  $-125.4$     &~$-119.48$ \\
      &  C   &                   &                   &               &~$-41.63$  \\ 
      &  T   &                   &                   &               &~$-77.85$  \\
      &Radius&  1.50(1)          & 1.50(4)           & 1.49          &~1.50      \\ \hline
\noalign{\hrule height 0.5pt}
\end{tabular}
\end{center}
\end{table}
%%%%%%%%%%%%%%%%%%%%%%%%%%%%%%

Next, we discuss the results of TOAMD with the bare AV6 $NN$ potential for $^3$H and $^4$He. 
The range parameters $\nu$ of $\Phi_{\rm AMD}$ were determined as 0.11 fm$^{-2}$ for $^3$H and 0.22 fm$^{-2}$ for $^4$He, respectively.
Similar to the case of the AV8$^\prime$ potential \cite{myo17a}, we obtain $\vec D_i=0$ for all the nucleons in $\Phi_{\rm AMD}$ of the two nuclei, which indicates the $s$--wave configurations.

In Table \ref{tab:AV6}, we show the results of TOAMD by adding each correlation term successively.
In each calculation, the range parameter $\nu$ is fixed, but the correlation functions are optimized.
We use the simple labels of D and S to express the correlation functions $F_D$ and $F_S$, respectively. 
The symbol +S is the result with the wave function $(1+F_S)\times \Phi_{\rm AMD}$.
The symbol +DD is the overall calculation of TOAMD up to the $F_D F_D$ term in Eq. (\ref{eq:TOAMD2}). 
The components of $F_S F_D$ and $F_D F_S$ are combined together in the results and denoted shortly as +DS.
The radii of $^3$H and $^4$He are 1.76 fm and 1.50 fm, respectively.

In Table \ref{tab:AV6_2}, we list the Hamiltonian components for two nuclei in TOAMD in comparison with other wave functions.
It was confirmed that the energies of $^3$H and $^4$He converged to the results of few-body calculations.
It is noted, in $^4$He, the kinetic energy in TOAMD shows somewhat smaller values than those of other wave functions, 
indicating the necessity of more correlations to increase the high-momentum components in TOAMD.

\begin{figure}[t]
\centering
\includegraphics[width=8.5cm,clip]{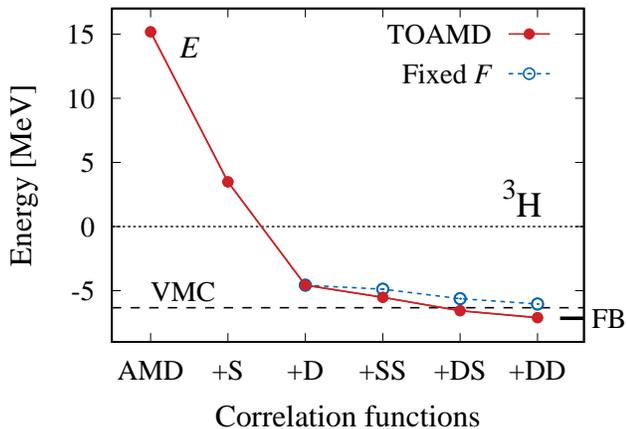}
\vspace{0.5mm}
\caption{Energy convergence of $^3$H using AV6 by adding each term of TOAMD successively with solid circles.
Open circles denoted as ``Fixed $F$'' indicate the results obtained using the fixed correlation functions in TOAMD. 
Dashed line and short solid-line represent the results of VMC and few-body (FB) calculations, respectively.}
\label{fig:ene_3H}
\end{figure}

In Fig.\,\ref{fig:ene_3H}, we show the total energy $E$ of $^3$H, obtained by adding the correlation terms in TOAMD successively.
We confirmed the converging behavior of the total-energy curve toward the values of few-body calculations.
It was found that the energy of TOAMD is lower than that of VMC.
This indicates that the variational accuracy of TOAMD is beyond that of VMC in the bare $NN$ interaction, similar to the central interaction case.

%%%%%%%%%%%%%%%%%%%%%%%%%%%%%%%%%%%%
%%%%%%%%%%%%%%%%%%%%%%%%%%%%%%%%%%%%
%%%%%%%%%%%%%%%%%%%%%%%%%%%%%%%%%%%%
\begin{figure}[t]
\centering
\includegraphics[width=8.5cm,clip]{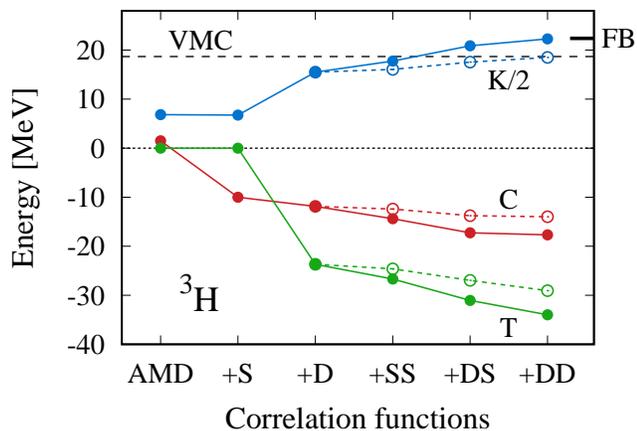}
\vspace{0.5mm}
\caption{Hamiltonian components of $^3$H using AV6 in TOAMD with solid circles.
The symbols K/2, C, and T indicate a half value of the kinetic energy, the central and tensor forces, respectively.
Open circles indicate the results using the fixed correlation functions (Fixed $F$) in TOAMD. 
Dashed line and short-solid line represent the half value of the kinetic energies in the calculations of VMC and few-body (FB) methods, respectively.}
\label{fig:ham_3H}
\end{figure}
%%%%%%%%%%%%%%%%%%%%%%%%%%%%%%%%%%%%

In TOAMD, all the correlation functions are independently optimized in every term in Eq.~(\ref{eq:TOAMD2}), which is a feature different from that of the Jastrow correlation method.
It is meaningful to investigate the effect of independent optimization of the correlation functions on the solutions of TOAMD.
We performed the following calculation:
First, $F_S$ and $F_D$ were determined in TOAMD truncated with $(1+F_S+F_D) \times \Phi_{\rm AMD}$.
Second, keeping the forms of $F_S$ and $F_D$, we performed the full calculation of TOAMD,
where only the weights of the seven components in Eq.~(\ref{eq:TOAMD2}) are variational parameters.
We show these constrained results denoted as ``Fixed $F$'' for $^3$H using open circles in Fig.\,\ref{fig:ene_3H}.
We can confirm the energy difference from the original TOAMD calculation.
This constraint finally provides the energy of $^3$H as $-6.04$ MeV with an energy loss of 1.06 MeV.
This energy is also close to that of VMC, which was $-6.33$ MeV. This property is interesting and reasonable
from the viewpoint of using common correlation functions in every pair.
From these results, it is clear that TOAMD is able to treat the $NN$ interaction using $F_D$ and $F_S$ better than VMC using the Jastrow correlation method.
Similarly, in Fig.\,\ref{fig:ham_3H}, we show the behavior of the Hamiltonian components of $^3$H.
We can confirm the differences between the values of TOAMD and the constrained case in every component.

%%%%%%%%%%%%%%%%%%%%%%%%%%%%%%%%%%%%
\begin{figure}[t]
\centering
\includegraphics[width=8.5cm,clip]{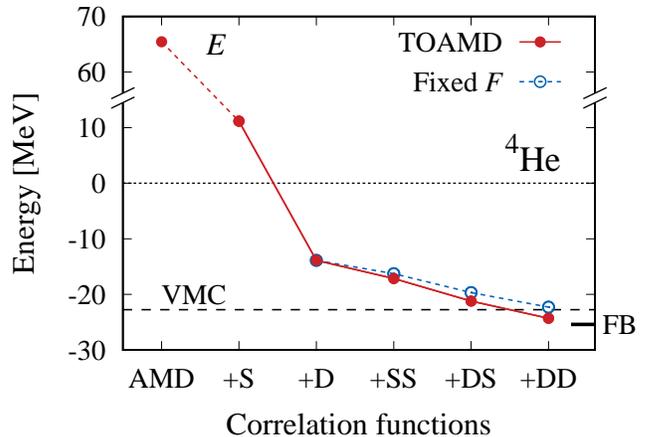}
\vspace{0.5mm}
\caption{Energy convergence of $^4$He using AV6.
Notations are the same as used in Fig. \ref{fig:ene_3H}.}
\label{fig:ene_4He}
\end{figure}
%%%%%%%%%%%%%%%%%%%%%%%%%%%%%%%%%%%%

%%%%%%%%%%%%%%%%%%%%%%%%%%%%%%%%%%%%
\begin{figure}[t]
\centering
\includegraphics[width=8.5cm,clip]{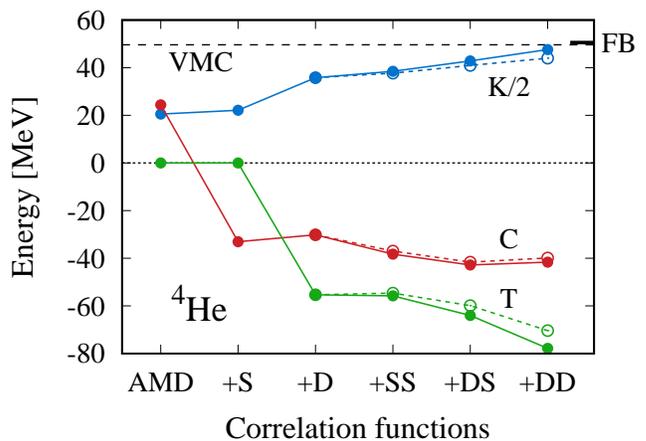}
\caption{Hamiltonian components of $^4$He using AV6.
Notations are the same as used in Fig. \ref{fig:ham_3H}.}
\label{fig:ham_4He}
\end{figure}
%%%%%%%%%%%%%%%%%%%%%%%%%%%%%%%%%%%%

Next, we discuss the case of $^4$He.
Figure.\,\ref{fig:ene_4He} shows the total energy obtained by adding the correlation function successively in TOAMD.
The behavior is very similar to that of $^3$H, and a good convergence with the correlation functions is observed. 
The energy of TOAMD is lower than that of VMC for $^4$He, similar to the case of $^3$H.
There exists slight deviations from the results of the few-body and GFMC calculations in Table \ref{tab:AV6_2}, which suggests the inclusion of the next order of TOAMD such as the triple products of correlation functions.

In the constrained ``Fixed $F$'' calculation for $^4$He, 
the energy is obtained as $-22.29$ MeV, with an energy loss of $2.02$ MeV.
This energy is also close to that of VMC ($-22.75$ MeV), similar to the $^3$H case.
Figure\,\ref{fig:ham_4He} explains the behavior of the Hamiltonian components of $^4$He showing the differences between the values of TOAMD and the constrained case in every component.
In particular, the central component provides the smaller difference than that of the tensor component in $^4$He in comparison with the $^3$H case in Fig.\,\ref{fig:ham_3H}.
This indicates the central correlations including the short-range correlation commonly act among nucleons for $^4$He, while this property is somewhat different for $^3$H.
This result comes from the fact that $^4$He is rigid because of the strong binding nature, which supports the Gaussian assumption of single-nucleon wave function in $\Phi_{\rm AMD}$ rather than the case of $^3$H.

For $^3$H and $^4$He, the calculation employing the common correlation functions of $F_D$ and $F_S$ in every term of TOAMD in Eq.~(\ref{eq:TOAMD2}) provides energies close to, but not below the values of VMC. 
This behavior seems reasonable, considering the treatment of correlation functions with common forms.
The present analysis shows that the correlation functions in each term of the power series in TOAMD should be different from each other. This property is favorable for the energy variation.
The flexible treatment of the correlation functions in TOAMD also contributes to the rapid energy convergence in the power series expansion.

TOAMD has an advantage to describe the clustering states on the basis of the AMD basis states.
As the next step, it is interesting to consider the cluster states such as $^8$Be consisting of two $^4$He.
In the TOAMD analysis, each $^4$He nucleus requires the double products of the correlation functions to obtain the converging solutions.
This fact suggests that the spatially developed two-$^4$He state will need at least fourth power of the correlation functions in TOAMD.
It is interesting to investigate description of the cluster states in TOAMD with the increase of the power of correlation functions.
There are several possibilities to reduce the computational time as to introduce the Monte-Carlo integral method.

%%%%%%%%%%%%%%%%%%%%%%%%%%%%%%%%%%%%%
\section{Summary}\label{sec:summary}
We developed a new variational method for the tensor--optimized antisymmetrized molecular dynamics (TOAMD).
In TOAMD, we introduce correlation functions to treat the tensor force and short-range repulsion in strong interactions.
We multiply these correlation functions with the basis states of the antisymmetrized molecular dynamics (AMD) in the form of a power series.
Each correlation function in each term of the multiple products is described independently.
This property of TOAMD is different from the ordinary Jastrow correlation method, where the correlation functions are multiplied to every pair once with a common form.

We have shown the results of $s$-shell nuclei using two kinds of nucleon--nucleon interaction including bare interaction in TOAMD within the double products of the correlation functions.
The energies and Hamiltonian components in TOAMD reproduce the few-body results of $^3$H and $^4$He satisfactorily.
The numerical accuracy of TOAMD is found to be beyond that of the variational Monte Carlo (VMC) calculation using the Jastrow correlation method.
The results indicate that the correlation functions should be different in each order of the power series expansion in TOAMD.  In a recent publication, there are more VMC results with AV18 and three-body interactions for $^{4}$He, $^{16}$O and $^{40}$Ca using the sophisticated Jastrow correlation method \cite{lonardoni17}. It is interesting to apply the TOAMD method to these cases.

We plan to increase the power of the multiple products of the correlation functions to the triple case in the expansion and apply TOAMD to the $p$-shell nuclei. 
We also plan to use three-nucleon interactions such as the Fujita-Miyazawa type,
which is treated similarly to that of many-body operators in the correlated Hamiltonian in TOAMD.

\section*{Acknowledgments}
This work was supported by JSPS KAKENHI Grants No. JP15K05091, No. JP15K17662, and No. JP16K05351.
Numerical calculations were partially performed on a computer system at RCNP, Osaka University.

\section*{References}
%\vfill\pagebreak
%%%%%%%%%%%%%%%%%%%%%%%%%%%%%%%%%%%%%%%%%%%%%%%%%%%%%%%%%%%%%
%%%%%%%%%%%%%%%%%%%%%%%%%%%%%%%%%%%%%%%%%%%%%%%%%%%%%%%%%%%%%
\def\JL#1#2#3#4{ {{\rm #1}} \textbf{#2}, #3 (#4)}  % Physical Review
\nc{\PR}[3]     {\JL{Phys. Rev.}{#1}{#2}{#3}}
\nc{\PRC}[3]    {\JL{Phys. Rev.~C}{#1}{#2}{#3}}
\nc{\PRA}[3]    {\JL{Phys. Rev.~A}{#1}{#2}{#3}}
\nc{\PRL}[3]    {\JL{Phys. Rev. Lett.}{#1}{#2}{#3}}
\nc{\NP}[3]     {\JL{Nucl. Phys.}{#1}{#2}{#3}}
\nc{\NPA}[3]    {\JL{Nucl. Phys.}{A#1}{#2}{#3}}
\nc{\PL}[3]     {\JL{Phys. Lett.}{#1}{#2}{#3}}
\nc{\PLB}[3]    {\JL{Phys. Lett.~B}{#1}{#2}{#3}}
\nc{\PTP}[3]    {\JL{Prog. Theor. Phys.}{#1}{#2}{#3}}
\nc{\PTPS}[3]   {\JL{Prog. Theor. Phys. Suppl.}{#1}{#2}{#3}}
\nc{\PTEP}[3]   {\JL{Prog. Theor. Exp. Phys.}{#1}{#2}{#3}}
\nc{\PRep}[3]   {\JL{Phys. Rep.}{#1}{#2}{#3}}
\nc{\PPNP}[3]   {\JL{Prog.\ Part.\ Nucl.\ Phys.}{#1}{#2}{#3}}
\nc{\JPG}[3]     {\JL{J. of Phys. G}{#1}{#2}{#3}}
\nc{\andvol}[3] {{\it ibid.}\JL{}{#1}{#2}{#3}}
%%%%%%%%%%%%%%%%%%%%%%%%%%%%%%%%%%%%%%%%%%%%%%%%%%%%%%%%%%%%%

\end{document}